\documentclass[aps,prb,twocolumn,amsmath,nofootinbib]{revtex4-1}

\usepackage{epsfig}
\usepackage[colorlinks,linkcolor=blue,anchorcolor=blue,
            citecolor=blue,urlcolor=blue,
            breaklinks=true]{hyperref}
\usepackage{graphics}
\usepackage{color}
\usepackage{dcolumn}
\usepackage{bm}
\usepackage{cases}
\usepackage{appendix}
\usepackage{lineno,hyperref}
\usepackage{ulem}

\newcommand{\tb}{\textbf}

\begin{document}
\author{Yong-Long Wang$^{1,2}$}
 \email{Email: wangyl@chenwang.nju.edu.cn}
\author{Guo-Hua Liang$^{1}$}
\author{Hua Jiang$^{1,2}$}
\author{Wei-Tao Lu$^{2}$}
\author{Hong-Shi Zong$^{1,3,4}$}
\email{Email: zonghs@nju.edu.cn}
\address{$^{1}$ Department of Physics, Nanjing University, Nanjing 210093, P. R. China}
\address{$^{2}$ School of Science and Institute of Condensed Matter Physics, Linyi University, Linyi 276005, P. R. China}
\address{$^{3}$ Joint Center for Particle, Nuclear Physics and Cosmology, Nanjing 210093, P. R. China}
\address{$^{4}$ State Key Laboratory of Theoretical Physics, Institute of Theoretical Physics, CAS, Beijing 100190, P. R. China}

\title{Transmission gaps from corrugations}
\begin{abstract}
A model including a periodically corrugated thin layer with GaAs substrate is employed to investigate the effects of the corrugations on the transmission probability of the nanostructure. We find that transmission gaps and resonant tunneling domains emerge from the corrugations, in the tunneling domains the tunneling peaks and valleys result from the boundaries between adjacent regions in which electron has different effective masses, and can be slightly modified by the layer thickness. These results can provide an access to design a curvature-tunable filter.
\bigskip

\end{abstract}
\maketitle

\section{INTRODUCTION}
With the advent and development of nanostructure technology, a variety of nanostructures with complex geometries were successfully fabricated, such as corrugated semiconductor films~\cite{Prinz2001, Prinz2006, Li2008, Turner2010, Mutilin2014, Zhang2015}, rolled-up nanotubes~\cite{Schmidt2001,Prinz2007, LiuF2007, Naumova2009, Froeter2013, Dastjerdi2015}, M\"{o}bius stripes~\cite{Tanda2002, Yoon2009, Monceau2012}, peanut-shaped $C_{60}$ polymers~ \cite{Onoe2003, Onoe2007, Onoe2008, Onoe2012, Onoe2015}. These successes in experiment found the basis of the emerging nanoelectronics. In the two-dimensional (2D) reduced systems, the dynamics of confined electron is affected by the surface curvature. As an important consequence, the geometric potential induced by the curvature appears in the effective surface quantum equation~\cite{Ferrari2008, BJensen2009a, Ortix2011, Wang2014} by the thin-layer quantization procedure~\cite{HJensen1971, Costa1981, Costa1982, Wang2016}. In the procedure, a squeezing potential is introduced to accomplish the reduction of the number of the spatial variables of the curved systems. Physically, the squeezing process is probably severe and breaks with the natural limits set by the uncertainty principle. Even so, the present quantization scheme can be safely applied to study the motion of confined electron when the quantum excitation energies in the normal direction are raised far beyond those in the tangential direction. Actually, the thin-layer quantization method has successfully been employed to calculate the band-structure of real systems~\cite{Aoki2001, Fujita2005, Koshino2005}, determine the localized surface states in geometrically deformed quantum systems~\cite{Goldstone1992, Cantele2000, Encinosa2003, Taira2007a, Taira2007b, BJensen2009b, Ortix2010}, and study the transport properties of electron confined in the systems with complex geometries~\cite{Marchi2005, Zhang2007, Cuoghi2009, Shima2009}. Furthermore, the experimental evidences for the geometrical effects of the curved surface have been presented, such as the realization of an optical analog of the curvature-induced geometric potential~ \cite{Szameit2010}, the observation of the influence of geometry on proximity effect~\cite{Kim2012} and the observation of Riemannian geometrical effects on electronic states~\cite{Onoe2012}. In other words, the geometrical deformation can be concluded as the presence of geometric potential in the dimensionally reduced quantum equation.

In the same vein, the advent and development of nanostructures have clearly led to a frontier field in 2D semiconductor research~\cite{Chang1992, Buot1993, Kushwaha1994, Lamberti2004, Jiang2012, XuS2015}. It is now possible to design and fabricate materials with prescribed electronic and photonic properties by artificial band-gap engineering. In order to satisfy the requirement of the development of nanoelectronics, theoretical physicists have tried to study the effects of the geometrical deformation on the tunneling rate~ \cite{Encinosa2000}, the electrical resistivity~ \cite{Ono2009, Ono2010a} and the persistent current~ \cite{Taira2010, Shima2012}. Moreover, on the basis of the geometric potential, quantum-electromechanical circuits~\cite{Blencowe2004, Chaplik2004, Xiang2013} and thin film transistors~\cite{Amalraj2014} have been proposed.

In fact, over the years, the electronic properties of the periodically curved surface have been the subject of active studies, both theoretical and experimental, due to the demand in understanding the physics involved and its great application potential. The ability to confine electron nearly to 2D regions on nanostructures has given a new impetus toward understanding the physics of reduced dimensionality systems. An important property of nanostructures is how their geometries affect their behaviors. Usually, one thinks of geometrical effects being connected primarily with the physical size and barrier configuration that define the confining region of the nanostructure.

\begin{figure}[htbp]
\centering
\includegraphics[width=0.3\textwidth]{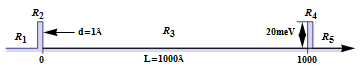}
\caption{\footnotesize (Color online) Model nanostructure schematic configuration parameters used to investigate the effects of the layer curvature and thickness on transmission probability.}\label{Model1}
\end{figure}

In the present study, we consider a model~ \cite{Encinosa2000} (shown in Fig.~\ref{Model1}) which contains two barriers and a well, the well is fabricated as a periodically corrugated thin layer. In the model, $R_1$ denotes free electron beam source, $R_2$ is a barrier with width $d=1{\AA}$, $R_3$ is a corrugated thin layer with width $L=1000{\AA}$, $R_4$ is the other barrier with width $d=1{\AA}$, and $R_5$ denotes a drain, from left to right.
The double barriers resonant tunneling structure is a typical microstructure that has been the focus of many investigations~\cite{Li1990, Chen1991a, Chen1991b, Kane1992, Doering1992, Nguyen2013}. In the present model, the corrugations are employed to provide transmission gaps and resonant tunneling domains, the double barriers and the boundaries between adjacent regions, in which electron has different effective masses, are adopted to generate the tunneling peaks and valleys in the tunneling domains. The theoretical basis is that when the spatial dimension is reduced to a scale being comparable with the de Broglie wavelength of electron in the vicinity of Fermi energy in the model, the wave nature of electron is expected to play an increasingly important role in transport properties.

This paper is organized as follows. In Sec.~\ref{2}, we briefly review the effective quantum equation for electron confined in a periodically corrugated layer by the thin-layer quantization scheme, and analyze the curvature-induced geometric potential. In Sec.~\ref{3}, we obtain a new geometric potential by the extended thin-layer quantization scheme, and also analyze the thickness-modified geometric potential. In Sec.~\ref{4}, we investigate the effects of the corrugations, double barriers and boundaries on transmission probability. Finally in Sec.~\ref{5} the conclusions are given.

\section{Quantum dynamics of a particle confined on a periodically corrugated surface}\label{2}
A 2D curved surface $S$ embedded in the usual three-dimensional (3D) space can be parametrized by $\vec{\tb{r}}=\vec{\tb{r}}(q_1,q_2)$, where $q_1$ and $q_2$ are the curvilinear coordinate variables over $S$. With respect to $q_1$ and $q_2$, two unit basis vectors $\vec{\tb{e}}_1$ and $\vec{\tb{e}}_2$ over $S$ are defined by $\vec{\tb{e}}_1=\frac{\partial\vec{\tb{r}}}{\partial q^1}/|\frac{\partial\vec{\tb{r}}}{\partial q^1}|$ and $\vec{\tb{e}}_2=\frac{\partial\vec{\tb{r}}}{\partial q^2}/|\frac{\partial\vec{\tb{r}}}{\partial q^2}|$, respectively. By introducing a curvilinear coordinate variable $q_3$ along the direction normal to $S$, a 3D subspace $V_N$ consisting of points near to $S$ and on $S$ can be described by $\vec{\tb{R}}(q_1,q_2,q_3)=\vec{\tb{r}}(q_1,q_2) +q_3\vec{\tb{e}}_n(q_1,q_2)$, where $\vec{\tb{e}}_n(q_1,q_2)$ is the unit basis vector perpendicular to $S$, with the definition $\vec{\tb{e}}_n=( \vec{\tb{e}}_1\times\vec{\tb{e}}_2)/ (|\vec{\tb{e}}_1\times\vec{\tb{e}}_2|)$.
In $V_N$, the covariant components of the metric tensor are defined by $G_{ij}=\frac{\partial\vec{\tb{R}}}{\partial q^i}\cdot \frac{\partial\vec{\tb{R}}}{\partial q^j}$ $(i,j=1,2,3)$. On $S$, the covariant components of the reduced metric tensor are determined by $g_{ab}=\frac{\partial\vec{\tb{r}}}{\partial q^a}\cdot \frac{\partial\vec{\tb{r}}}{\partial q^b}$ $(a,b=1,2)$. The relationships between $G_{ab}$ and $g_{ab}$ are
\begin{equation}\label{RelationVnS}
G_{ab}=g_{ab}+(\alpha g+g^T\alpha^T)_{ab}q_3+(\alpha g\alpha^T)_{ab}(q_3)^2,
\end{equation}
and $G_{a3}=G_{3a}=0$, $G_{33}=1$, where $T$ denotes the matrix transpose, $\alpha$ is the Weingarten curvature matrix
\begin{equation}\label{Alpha0}
\alpha=\frac{1}{g}
\left (
\begin{array}{cc}
g_{12}h_{21}-g_{22}h_{11} & g_{21}h_{11}-g_{11}h_{21}\\
g_{12}h_{22}-g_{22}h_{12} & g_{12}h_{21}-g_{11}h_{22}
\end{array}
\right ),
\end{equation}
wherein $h_{ab}$ are the coefficients of the second fundamental form, $h_{ab}=\vec{\tb{e}}_n\cdot\frac{\partial^2\vec{\tb{r}}}{\partial q^a\partial q^b}$. By means of $\alpha$, the mean curvature $M$ is $M=\frac{1}{2}\mathrm{Tr}(\alpha)$ and the Gaussian curvature $K=\det(\alpha)$, and then the relation between $G=\det(G_{ij})$ and $g=\det(g_{ab})$ is $G=f^2g$ with $f=1+2Mq_3+K(q_3)^2$.

Basing on the above mathematical formula, we can confine a free electron on a curved surface by the thin-layer quantization scheme~\cite{HJensen1971, Costa1981}, the effective surface Schr\"{o}dinger equation is obtained as
\begin{equation}\label{SSE0}
-\frac{\hbar^2}{2m} \frac{1}{\sqrt{g}}\partial_a(\sqrt{g}g^{ab}\partial_b \chi_s)+V_g\chi_s=E_s\chi_s,
\end{equation}
where $V_g$ is the geometric potential
\begin{equation}\label{GP0}
V_g=-\frac{\hbar^2}{2m}(M^2-K).
\end{equation}

\begin{figure}[htbp]
  \centering
  \includegraphics[width=0.27\textwidth]{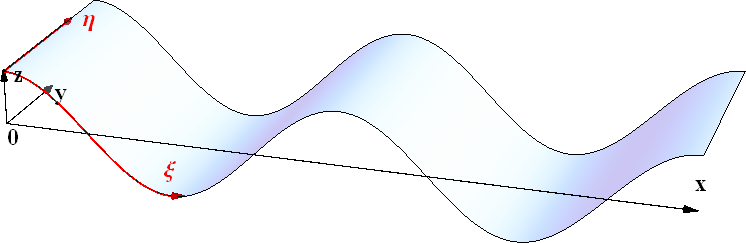}
  \caption{\footnotesize (Color online) Schematic of a periodically corrugated surface described by $z=a\cos(\gamma x)$. Here $a$ and $2\pi/\gamma$ are the amplitude and period length of the corrugations, respectively. $(\xi,\eta)$ denotes the two curvilinear coordinates over surface.}\label{surface}
\end{figure}

Practically, nanocorrugated thin-films are often found in nanomaterial experiments. For the sake of simplicity, a curved surface $\mathcal{S}$ shown in Fig.~\ref{surface} is considered. It is corrugated along the direction of $x$ with period $2\pi/\gamma$, amplitude $a$, but is flat along that of $y$. In the Monge form, $\mathcal{S}$ can be described as
\begin{equation}\label{SurfaceM}
\vec{\tb{r}}=\vec{\tb{e}}_x x+\vec{\tb{e}}_y y+\vec{\tb{e}}_z a\cos(\gamma x).
\end{equation}

According to the procedure mentioned above, from Eq.~\eqref{SurfaceM} we obtain three unit basis vectors and corresponding three derivative elements
\begin{equation}\label{VectorOneform}
\begin{split}
&\vec{\tb{e}}_{\xi}=w\vec{\tb{e}}_x+w a\gamma\sin(\gamma x)\vec{\tb{e}}_z, \quad\quad d\xi=\frac{1}{w}dx,\\
&\vec{\tb{e}}_{\eta}=\vec{\tb{e}}_y, \quad\quad\quad\quad\quad\quad \quad\quad\quad\quad d\eta=dy,\\
&\vec{\tb{e}}_n=-w a\gamma\sin(\gamma x)\vec{\tb{e}}_x+w\vec{\tb{e}}_z, \quad dq_3=dq_3,
\end{split}
\end{equation}
respectively, with
\begin{equation}\label{Factorw}
 w=\frac{1}{\sqrt{1+a^2\gamma^2\sin^2(\gamma x)}}.
\end{equation}
Subsequently, we obtain $(G_{ab})$ and $G$
\begin{equation}\label{3DMT}
(G_{ab})=
\left (
\begin{array}{cc}
(1+q_3k)^2 & 0\\
0 & 1\\
\end{array}
\right ),\quad G=(1+q_3k)^2,
\end{equation}
the Weigarten curvature matrix $\alpha$
\begin{equation}\label{Alpha}
\alpha=
\left (
\begin{array}{cc}
k & 0\\
0 & 0
\end{array}
\right ),
\end{equation}
$(g_{ab})$ and $g$
\begin{equation}\label{2DMT}
(g_{ab})=
\left (
\begin{array}{cc}
1 & 0 \\
0 & 1
\end{array}
\right ), \quad g=1,
\end{equation}
where $k=w^3a\gamma^2\cos(\gamma x)$. In the above calculated process, the factor $f$, the mean curvature $M$ and the Gaussian curvature $K$ can be given as
\begin{equation}\label{Factorf}
f=1+kq_3,\quad M=\frac{1}{2}k,\quad K=0.
\end{equation}
Consequently, the expected Schr\"{o}dinger equation~\eqref{SSE0} is
\begin{equation}\label{SSE1}
-\frac{\hbar^2}{2m^*}[w\frac{\partial}{\partial x}(w\frac{\partial}{\partial x})+ \frac{\partial^2}{\partial y^2}]\chi_s+V_g\chi_s =E_s\chi_s,
\end{equation}
where $m^*$ denotes the effective mass of electron, and $V_g$ is the geometric potential
\begin{equation}\label{GP1}
V_g=-\frac{\hbar^2}{8m^*}\frac{[a\gamma^2\cos(\gamma x)]^2}{[1+a^2\gamma^2\sin^2(\gamma x)]^3}.
\end{equation}
These results are the same as those given by S. Ono and H. Shima~\cite{Ono2009}. The quantum motion in the normal direction is neglected, because the introduced squeezing potential raises the quantum excitation energies in normal direction far beyond those in tangential direction~\cite{HJensen1971, Costa1981}.

\begin{figure}[htbp]
  \centering
  \includegraphics[width=0.44\textwidth]{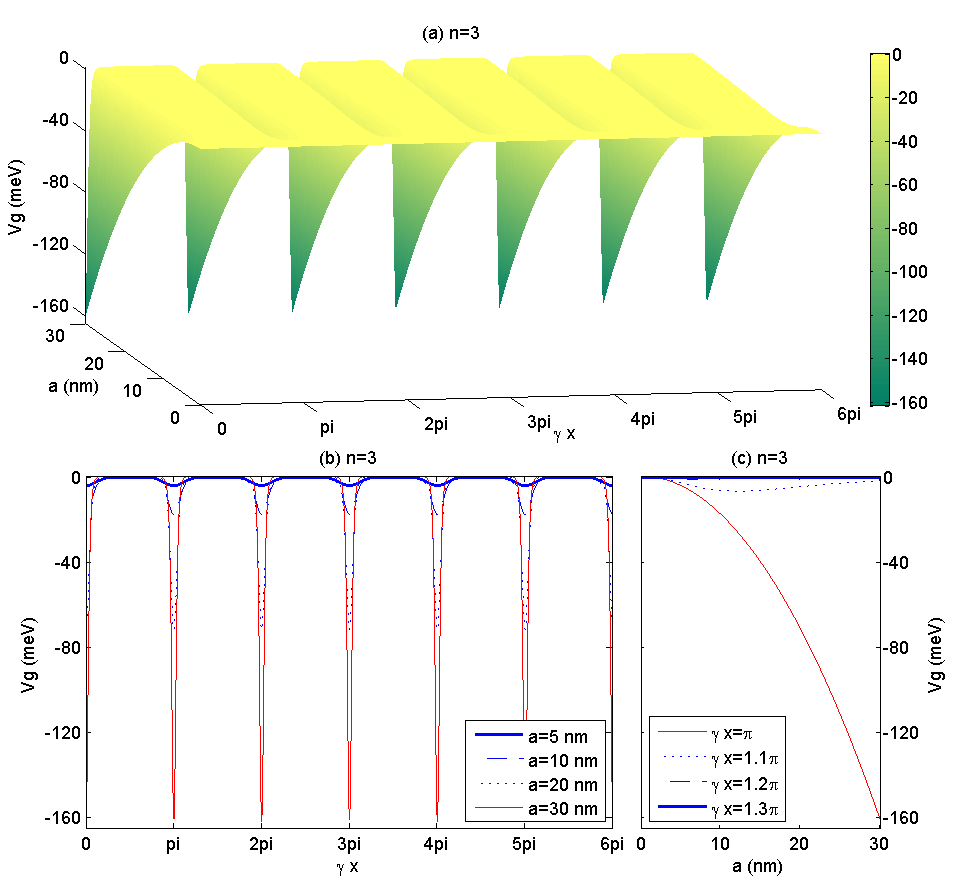}\\
  \caption{\footnotesize (Color online) (a) Surface of $V_g$ with $n=3$, as a function of $\gamma x$ and $a$. (b) Slices of the geometric potential $V_g$ at $a=5nm, 10nm, 20nm, 30nm$, as a function of $\gamma x$. (c) Slices of the geometric potential $V_g$ at $\gamma x=\pi, 1.1\pi, 1.2\pi, 1.3\pi$, as a function of $a$.}\label{GPFig2}
\end{figure}

As the central result of the thin-layer quantization scheme, the geometric potential appears in the expectant quantum equation. With $n=3$ and $m^*=0.067m_0$ ($0.067m_0$ is the effective mass of electron in GaAs substrate) the geometric potential $V_g(x)$ shown in Fig.~\ref{GPFig2} is a function of $\gamma x$ and $a$, where $n$ denotes the period number of the corrugations in $R_3$, $m_0$ is the mass of a free electron. As observed in Fig.~\ref{GPFig2} (a), the $x$ dependence of $V_g(x)$ deviates considerably from a cosinusoidal surface, whereas the surface corrugation is exactly cosinusoidal. It is shown in Fig.~\ref{GPFig2} (b) that the downward peaks are formed at $\gamma x=\pm l\pi (l=0,1,2,\cdots)$, here the height of the surface $\mathcal{S}$ is either maximum $(z=a)$ or minimum $(z=-a)$. The amplitude of the peaks grows sharply with increasing $a$ that is highlighted in Figs.~\ref{GPFig2} (b) and (c). As a consequence, we can provide a list of attractive potential wells by introducing corrugations, tune their depths by $a$, and design their number by $n$.
\section{The modification of the layer thickness to the geometric potential}\label{3}
\begin{figure}[htbp]
  \centering
  \includegraphics[width=0.27\textwidth]{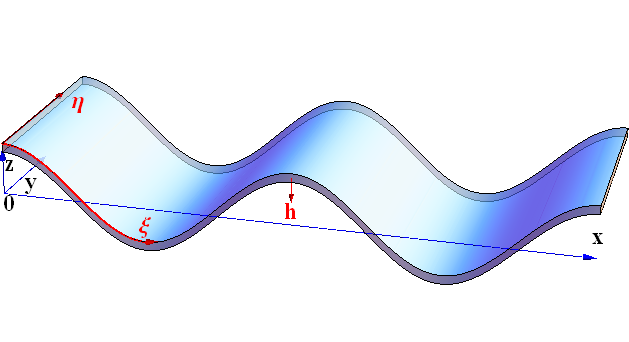}\\
  \caption{\footnotesize (Color online) Schematic of a periodically curved layer with a uniform thickness $h$. The standard surface of the layer is described by $z=a\cos(\gamma x)$. Here $a$ and $2\pi/\gamma$ are the amplitude and period length of corrugations, respectively. $(\xi,\eta)$ are the two curvilinear coordinates over layer. }\label{Layer}
\end{figure}

Really, the corrugated layer has certain thickness. The thickness effects on the effective quantum equation~\eqref{SSE1} can be investigated by the extended thin-layer quantization scheme~\cite{Wang2016}. In the present model the thickness effects on the kinetic term can be neglected, because the thickness lengthens the displacement of electron across the layer very slightly, whereas a thickness-modified geometric potential $V_g^{\prime}$ can be given as
\begin{equation}\label{MGP1}
V^{\prime}_g=V_g+\frac{\hbar^2}{4m^*} \frac{[a\gamma^2\cos(\gamma x)]^3} {[1+a^2\gamma^2\sin^2(\gamma x)]^{\frac{9}{2}}}h,
\end{equation}
where $h$ denotes the layer thickness sketched in Fig.~\ref{Layer} and $V_g$ is the geometric potential~\eqref{GP1}. The second term on the right hand side of Eq.~\eqref{MGP1} attributes to the layer thickness. It needs to notice that $h$ must be less than the minimum curvature radius on $\mathcal{S}$. For an arbitrary point on $\mathcal{S}$, there is only one principle curvature $k=w^3a\gamma^2\cos(\gamma x)$, the corresponding curvature radius is
\begin{equation}\label{Radius}
\rho=\frac{[1+a^2\gamma^2\sin^2(\gamma x)]^{\frac{3}{2}}}{a\gamma^2\cos(\gamma x)}.
\end{equation}
It is obvious that the minimum curvature radius is $\rho_{min}=1/(a\gamma^2)$ at $\gamma x=n\pi$ $(n=0,\pm 1,\pm2,\cdots)$. When $h=\rho_{min}$, at new points (they have a uniform distance $\rho_{min}$ to $\mathcal{S}$) "$h-\rho_{min}$" is $0$, the corresponding mean curvature $M$ becomes $\infty$. In order to avoid the trouble, we primitively define $h$ ranging between $0$ and $1/(2a\gamma^2)$.

\begin{figure}[htbp]
  \centering
  \includegraphics[scale=0.33]{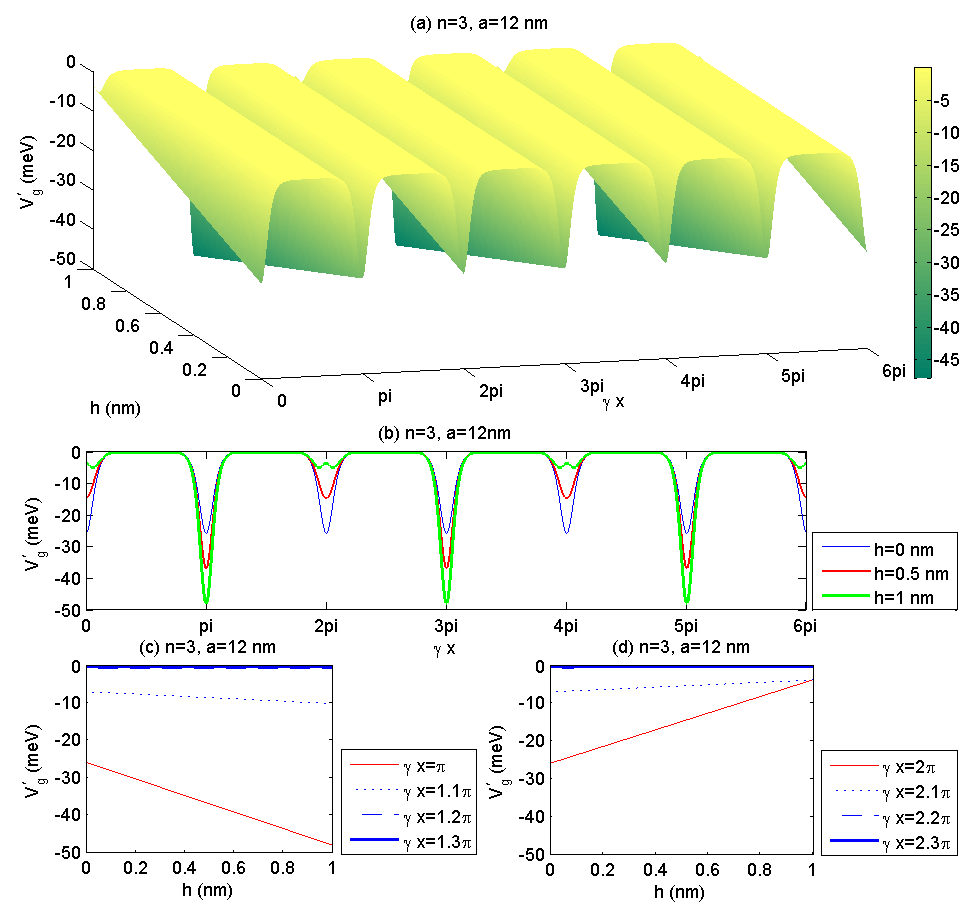}\\
  \caption{\footnotesize (Color online) (a) Surface of $V_g^{\prime}$ with $n=3$ and $a=12 nm$, as a function of $\gamma x$ and $h$. (b) Slices of $V_g^{\prime}$ at $h=0nm, 0.5nm, 1nm$, as a function of $\gamma x$ with $n=3$ and $a=12nm$. (c) Slices of $V_g^{\prime}$ at $\gamma x=\pi, 1.1\pi, 1.2\pi, 1.3\pi$, as a function of $h$ with $n=3$ and $a=12nm$. (d) Slices of $V_g^{\prime}$ at $\gamma x=2\pi, 2.1\pi, 2.2\pi, 2.3\pi$, as a function of $h$.}\label{MGPFig3}
\end{figure}

\begin{figure}[htbp]
  \centering
  \includegraphics[width=0.44\textwidth]{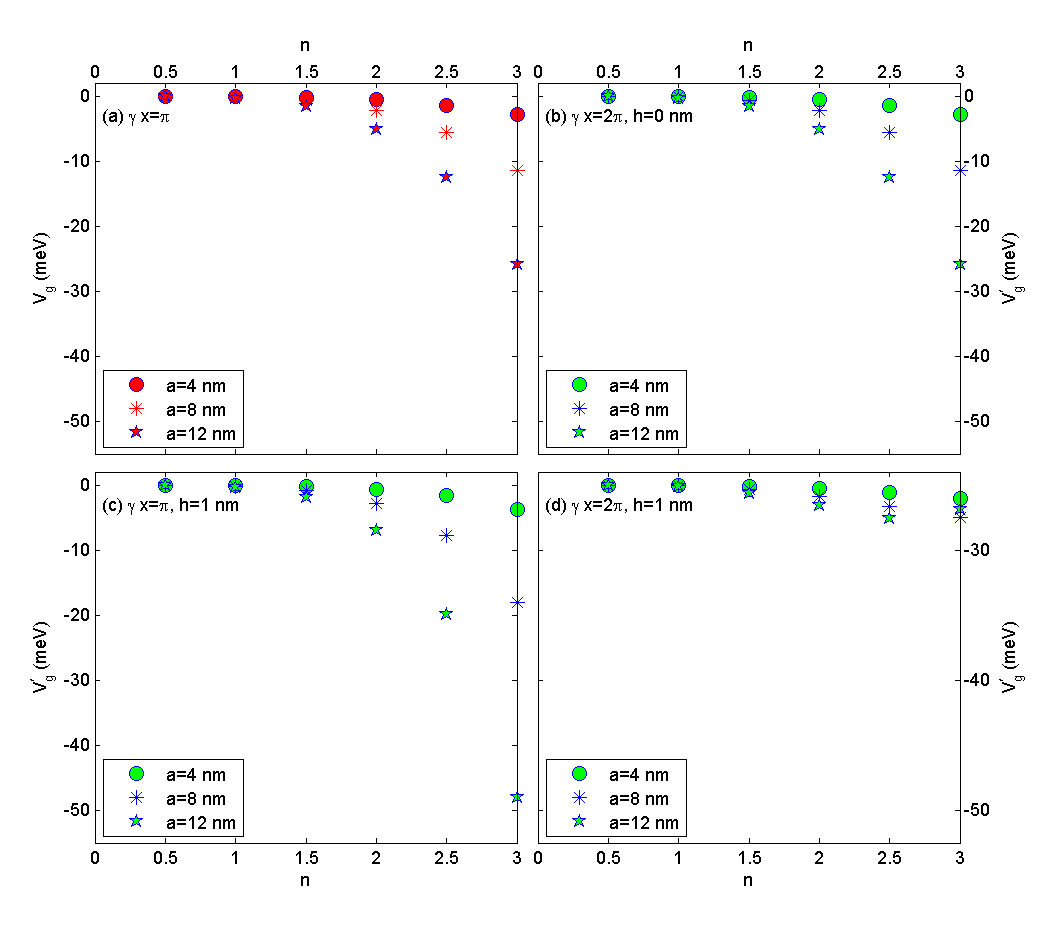}\\
  \caption{\footnotesize (Color online) (a) The downward peaks of $V_g$ at $\gamma x=\pi$ versus $n$ for $a=4 nm, 8 nm, 12 nm$. (b) The downward peaks of $V_g^{\prime}$ at $\gamma x=2\pi$ and $h=0 nm$ versus $n$ for $a=4 nm, 8 nm, 12 nm$. (c) The downward peaks of $V_g^{\prime}$ at $\gamma x=\pi$ and $h=1 nm$ versus $n$ for $a=4 nm, 8 nm, 12 nm$. (d) The downward peaks of $V_g^{\prime}$ at $\gamma x=2\pi$ and $h=1 nm$ versus $n$ for $a=4 nm, 8 nm, 12 nm$.}\label{NPoints}
\end{figure}

As an important result for the extended thin-layer quantization scheme~\cite{Wang2016}, the modification of the layer thickness is included in $V_g^{\prime}(x)$, as a function of $\gamma x$ and $h$, described in Fig.~\ref{MGPFig3} with $n=3$ and $a=12 nm$. As mentioned above, the downward peaks are still formed at $\gamma x=\pm l\pi (l=0, 1, 2, \cdots)$. However, there is a change shown in Fig.~\ref{MGPFig3} (a), the downward peaks at $\gamma x=\pm(2m+1)\pi$ $(m=0, 1, 2, \cdots)$ grow with increasing $h$, but those at $\gamma x=\pm 2m\pi$ dwarf, that is highlighted in Figs.~\ref{MGPFig3} (b), (c) and (d).

The scales of $V_g(x)$ and $V_g^{\prime}(x)$ are substantially influenced by $n$ described in Figs.~\ref{NPoints} (a) and (b), the larger the period number $n$, the deeper the peaks at $\gamma x=l\pi (l=0, 1, 2, \cdots)$. Obviously, as shown in Figs.~\ref{NPoints} (c) and (d), $h$ evidently deepens the peaks at $\gamma x=(2l+1)\pi$, but remarkably shallows those at $\gamma x=2l\pi$. In other words, we can adjust the discrepancy between adjacent wells by depositing the layer thickness $h$ finely.
\section{Transmission probability in the periodically corrugated thin layer}\label{4}
\subsection{Transmission probability}
According to Eq.~\eqref{SSE1}, with the limit $h\to 0$ the quantum equation for an electron in the model depicted in Fig.~\ref{Model1} can be
\begin{equation}\label{4SE1}
-\frac{\hbar^2}{2m^*}w\frac{d}{dx}[w\frac{d}{dx}\psi(x)] +V(x)\psi(x)=E\psi(x),
\end{equation}
where $m^*=0.067m_0$ (the effective mass of electron in GaAs substrate) if $x\in R_3$ and $m^*=m_0$ otherwise, $w=1/\sqrt{1+a^2\gamma^2\sin^2(\gamma x)}$ if $x\in R_3$ and $w=1$ otherwise, $\psi$ is a wave function, $E$ is the energy with respect to $\psi$ and $V(x)$ is
\begin{equation}\label{4Potential1}
V(x)=
\begin{cases}
& 0 meV, \quad x\in R_1,\\
& 20 meV, \quad x\in R_2,\\
& V_g(x), \quad x\in R_3,\\
& 20 meV, \quad x\in R_4,\\
& 0 meV, \quad x\in R_5,
\end{cases}
\end{equation}
wherein $V_g(x)$ is the geometric potential~\eqref{GP1} and $meV$ denotes milli electron volts. When the effects of the layer thickness are considered, $V_g(x)$ should be replaced by $V_g^{\prime}(x)$ in Eq.~\eqref{MGP1}.

With the help of the transfer matrix technique~ \cite{Ando1987}, instead of dealing with continuous variations of $V(x)$ in $R_3$, we split $R_3$ up into segments, in each segment $V(x)$ can be regarded as a constant. And then let us assume $R_3$ will be a sequence $N_3$ small segments, $R_2$ one segment ($N_2=1$) and $R_4$ one segment ($N_4=1$). It is straightforward to obtain that the total number of segments is $N=N_2+N_3+N_4$, that of boundaries is $N+1$. For an arbitrary segment, the $j$th region, in which the wave function $\psi_j(x)$ can be given by
\begin{equation}\label{4WaveFunction1}
\psi_j(x)=A_j\exp[ik_j \xi(x)]+B_j\exp[-ik_j \xi(x)],
\end{equation}
where
\begin{equation}\label{4WaveVector1}
k_j=\frac{\sqrt{[2m_j^*(E-U_j)]}}{\hbar},
\end{equation}
wherein $\hbar$ is the reduced Planck's constant, $m_j^*$ is the effective mass for electron at the middle point in the $j$th region, $U_j$ is a constant, $U_j=V(\frac{x_{j-1}+x_j}{2})$. For the wave function~ \eqref{4WaveFunction1}, it is particular that the curvilinear coordinate variable $\xi(x)$ as a function of $x$ has the derivative form $d\xi=\sqrt{1+a^2\gamma^2\sin^2(\gamma x)}dx$ if $x\in R_3$ and $d\xi=dx$ otherwise.

On account of the continuities of $\psi_j(x)$ and $\psi_j^{\prime}(x)/m_j^*$ at each boundary, we can determine $A_j$ and $B_j$ in Eq.~\eqref{4WaveFunction1} by the following multiplication
\begin{equation}\label{4TM1}
\left (
\begin{array}{c}
A_j\\
B_j
\end{array}
\right )=\prod_{l=0}^{j-1}M_l
\left (
\begin{array}{c}
A_0\\
B_0
\end{array}
\right ),
\end{equation}
where
\begin{widetext}
\begin{equation}\label{4TM2}
M_l=\frac{1}{2}
\left [
\begin{array}{cc}
(1+s_l)\exp[-i(k_{l+1}-k_l)x_l] & (1-s_l)\exp[-i(k_{l+1}+k_l)x_l]\\
(1-s_l)\exp[i(k_{l+1}+k_l)x_l] & (1+s_l)\exp[i(k_{l+1}-k_l)x_l]
\end{array}
\right ]
\end{equation}
\end{widetext}
with
\begin{equation}\label{4S}
s_l=\frac{m_{l+1}^*}{m_l^*}\frac{k_l}{k_{l+1}}.
\end{equation}
As $j=N$, the equation~\eqref{4TM1} becomes
\begin{equation}\label{4TM3}
\left (
\begin{array}{c}
A_N\\
B_N
\end{array}
\right )=M\left (
\begin{array}{c}
A_0\\
B_0
\end{array}
\right ),
\end{equation}
where
\begin{equation}\label{4TM4}
M=\left (
\begin{array}{cc}
M_{11} & M_{12} \\
M_{21} & M_{22}
\end{array}
\right )=
\prod_{l=1}^{N}M_l.
\end{equation}

Without any loss of generality, we assume that in the model under study $R_1$ is an electron source consisting of free electrons, and $R_5$ is a drain. Using the plane wave approximation, the wave function in $R_1$ can be
\begin{equation}\label{4WaveFunction2}
\psi_0(x)=\exp(ik_0x_0)+B_0\exp(-ik_0x_0),
\end{equation}
and in $R_5$
\begin{equation}\label{4WaveFunction3}
\psi_N(x)=A_N\exp(ik_Nx_N),
\end{equation}
where $B_0$ and $A_N$ are the coefficients of reflection and transmission, respectively. In the case of $A_0=1$, $B_N=0$, $m_0^*=m_{N+1}^*=m_0$ and $k_0=k_{N+1}$, we can obtain the transmission amplitude $A_N$ and the transmission probability $T$ as
\begin{equation}\label{4TransAmpl2}
A_N=\frac{1}{M_{22}},
\end{equation}
and
\begin{equation}\label{4TransProb2}
T=\frac{1}{(M_{22})^2},
\end{equation}
respectively.

\subsection{Numerical results and analysis}

\begin{figure}[htbp]
\includegraphics[width=0.44\textwidth]{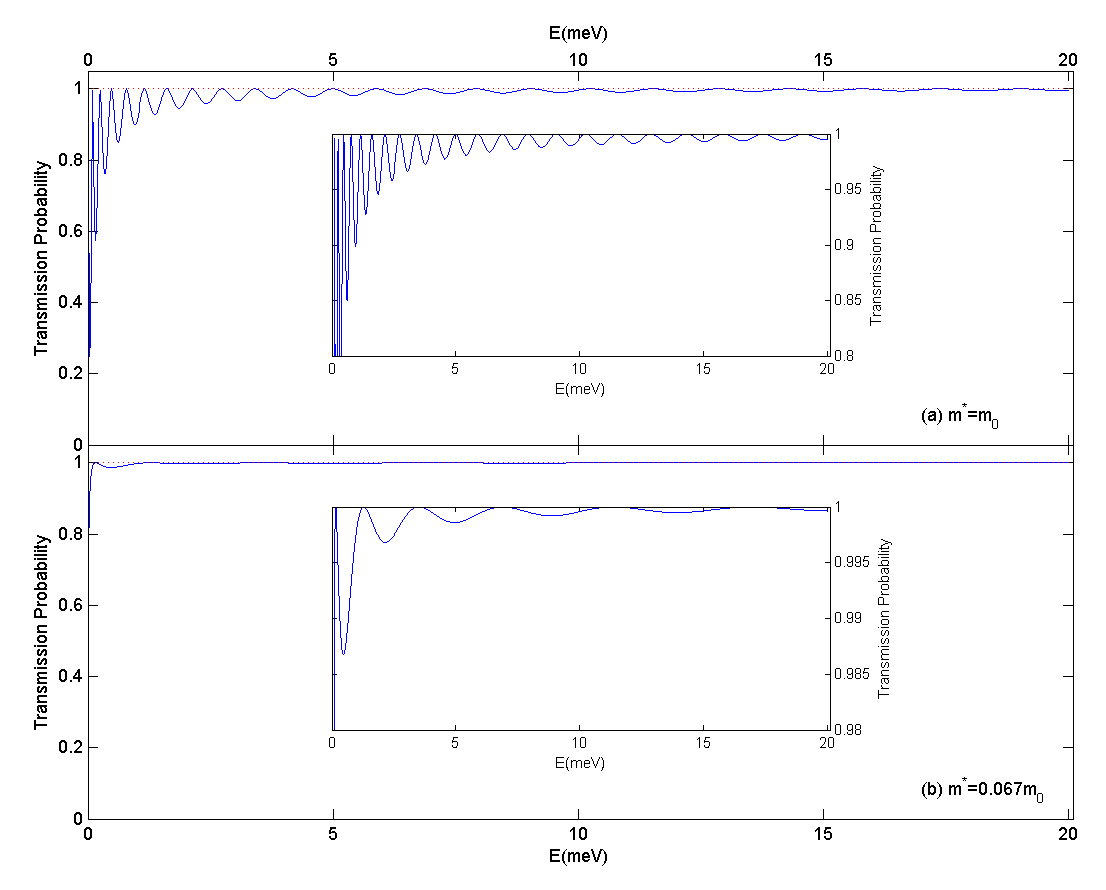}
\caption{\footnotesize (Color online) Transmission probability versus incident energy $E$ for the model with $V(x)=20meV$ in $R_2$ and $R_4$, $V(x)=0meV$ otherwise, and $m^*$ for (a) $m^*=m_0$ and (b) $m^*=0.067m_0$.}\label{Fig7}
\end{figure}

In this subsection, we will investigate how the corrugations affect the transmission probability by the transfer matrix method. Before beginning the investigation, we briefly analyze the effects of other components in the model on the transmission probability. The effective masses of electron in the model are closely related to the resonant tunneling peaks as shown in Figs.~\ref{Fig7}. When the effective mass is less, the number of the peaks is less, their amplitudes smaller. These results are highlighted in the insets in Fig.~\ref{Fig7}.

\begin{figure}[htbp]
\includegraphics[width=0.44\textwidth]{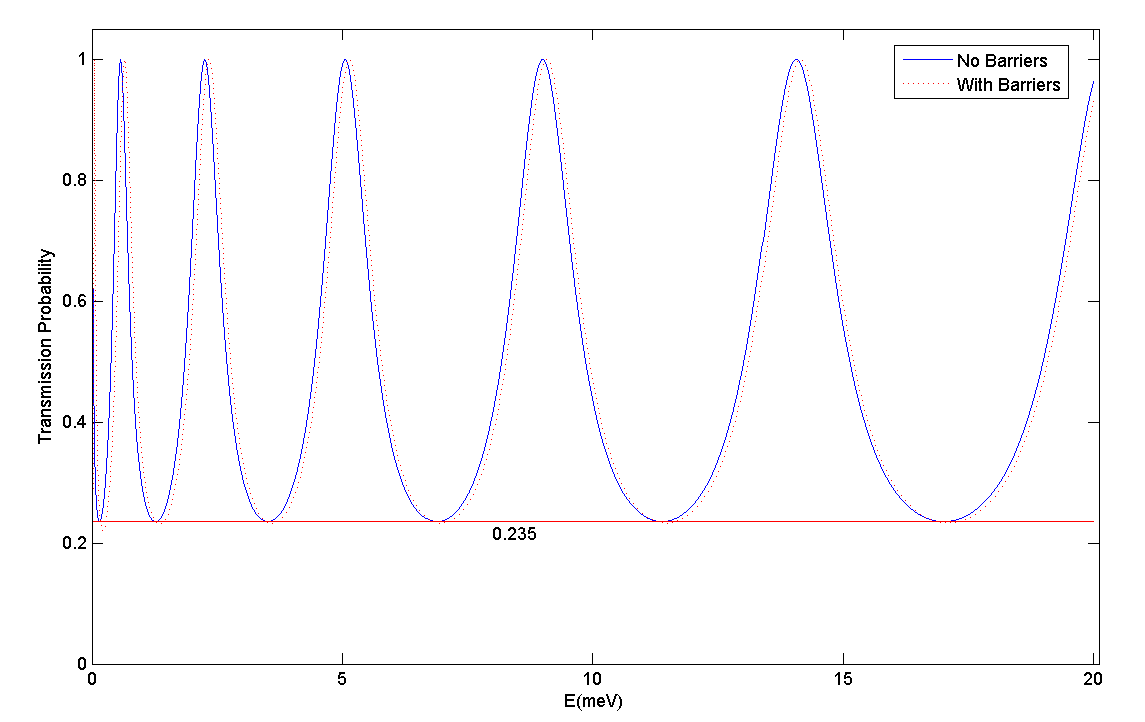}
\caption{\footnotesize (Color online) Transmission probability versus incident energy $E$ for the model with the effective mass $m^*=0.067m_0$ in $R_3$, $m^*=m_0$ otherwise, no barriers plotted as solid blue curve, with barriers done as dotted red curve.}\label{Fig8}
\end{figure}

In the case of the model with $m^*=0.067m_0$ in $R_3$ and $m^*=m_0$ otherwise, the two boundaries between $R_2$ and $R_3$, $R_3$ and $R_4$ are essential to influence the transmission probability. In response to the boundaries, the amplitudes of the tunneling peaks considerably grow shown in Fig.~\ref{Fig8}. In other words, these peaks are mostly provided by the continuities of $\psi(x)$ and $\psi^{\prime}(x)/m^*$, the contributions of the double barriers could be neglected. It is worth noticing that the peaks almost reach $1$, the valleys are about $0.235$. Physically, the peaks occur when the length of $R_3$  strictly equals integer multiple times of a half wave length of electron, the bottom of the valleys is eventually determined by the ratio of the effective masses in $R_3$ and $R_2$ (or $R_4$).

\begin{figure}[htbp]
\centering
\includegraphics[width=0.44\textwidth]{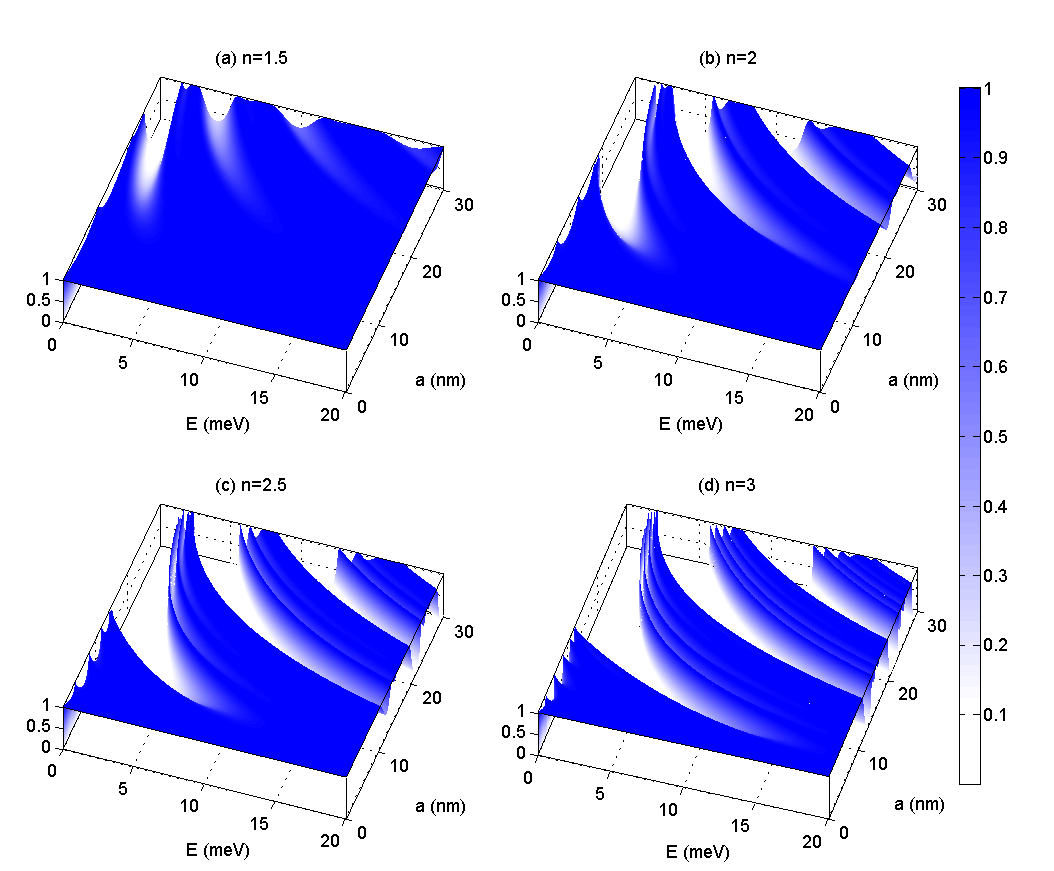}
\caption{\footnotesize (Color online) Surface plot of the transmission probability as a function of $E$ and $a$ at (a) $n=1.5$, (b) $n=2$, (c) $n=2.5$ and (d) $n=3$, with $m^*=0.067m_0$ for all regions, $V(x)=V_g(x)$ in $R_3$ and $V(x)=0meV$ otherwise.}\label{Fig9}
\end{figure}

In a special case, the model without the double barriers, with $m^*=0.067m_0$ in all regions, $V(x)=V_g(x)$ in $R_3$ and $V(x)=0 meV$ otherwise, transmission gaps~\cite{Khelif2003, Chen2009a, Chen2009b, Lu2012, XuY2015} and resonant tunneling domains~\cite{LiuXW1993, Guo1998, Zeng1999, Xu2014, Pham2015} appear in the present system shown in Fig.~\ref{Fig9}. The presence of the transmission gaps is the most fascinating finding in this study. It is readily proved that the widths of the transmission gaps grow with increasing the corrugation amplitude $a$. The cause is that the larger the distances between adjacent wells in the geometric potential $V_g(x)$, the less the communication or coupling between adjacent wells~\cite{LiuXW1994}. As periodical corrugations presented in $R_3$, the factor $w$ in Eq.~\eqref{4SE1} is a function of $x$ as $w=1/\sqrt{1+a^2\gamma^2\sin^2(\gamma x)}$, in $R_3$ $x_j$ must be replaced by $\xi_j$
\begin{equation}\label{Xi} \xi_j=\int_{0}^{x_j}\sqrt{1+a^2\gamma^2\sin^2(\gamma x)}dx.
\end{equation}
According to this integral, the distance between adjacent wells naturally grows with increasing the corrugation amplitude. Another aspect of the transmission probability is that the transmission gaps become wider when $n$ is larger with a fixed $a$, the tunneling domains do narrower correspondingly. The reason is that the more the number of the wells, the stronger their reflections. The resonant tunneling domains are formed essentially by the coupling between adjacent wells, with respect to resonant energy domains~\cite{LiuXW1994, Lu2013}. As an application potential, the transmission gaps mean that electron is mostly reflected, but the tunneling domains do that electron can readily pass.

\begin{figure}[htbp]
\includegraphics[width=0.44\textwidth]{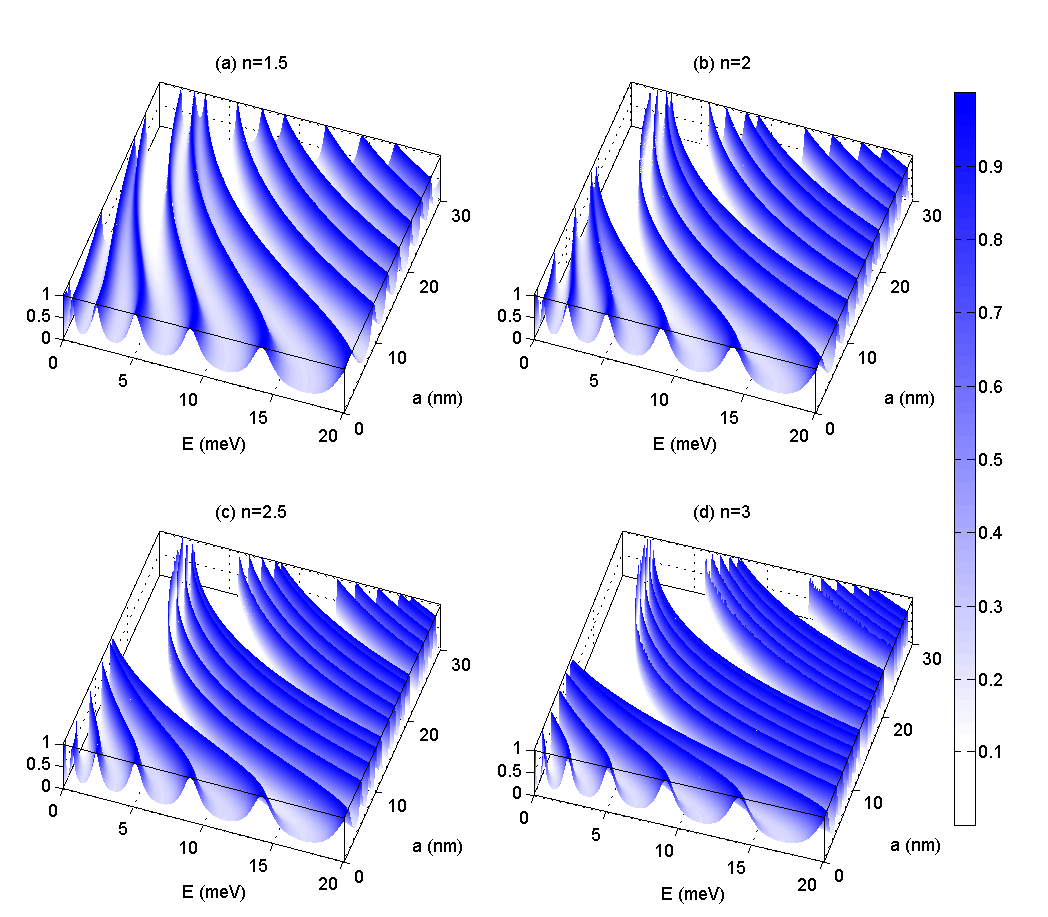}
\caption{\footnotesize (Color online) Surface plot of the transmission probability as a function of $E$ and $a$ at (a) $n=1.5$, (b) $n=2$, (c) $n=2.5$ and (d) $n=3$, with $m^*=0.067m_0$ in $R_3$ and $m^*=m_0$ elsewhere, $V(x)=V_g(x)$ in $R_3$, $V(x)=20 meV$ in $R_2$ and $R_4$ and $V(x)=0 meV$ otherwise.}\label{Fig10}
\end{figure}

In the considered case, the model with $m^*=0.067m_0$ in $R_3$ fabricated by GaAs substrate and $m^*=m_0$ otherwise, with $V(x)$ in Eq.~\eqref{4Potential1}, the transmission probability as a function of $E$ and $a$ is described in Fig.~\ref{Fig10} for (a) $n=1.5$, (b) $n=2$, (c) $n=2.5$ and (d) $n=3$. In striking contrast to Fig.~\ref{Fig9}, Fig.~\ref{Fig10} shows that tunneling peaks and valleys evidently occur in the tunneling domains. Significantly, the transmission gaps are still kept. That is to say that the boundaries and the double barriers cannot considerably influence the transmission gaps, but they construct the tunneling peaks and valleys in the tunneling domains, especially when the amplitude $a$ is small. This result can be directly manifested by that the bottom of the valleys is still a certain constant at $a=0$, which well agrees with that described in Fig.~\ref{Fig8}. Additionally, it is worthwhile to note that the number of the tunneling peaks in each of the tunneling domains is equivalent to that of the wells in the geometric potential.

\begin{figure}[htbp]
\includegraphics[width=0.44\textwidth]{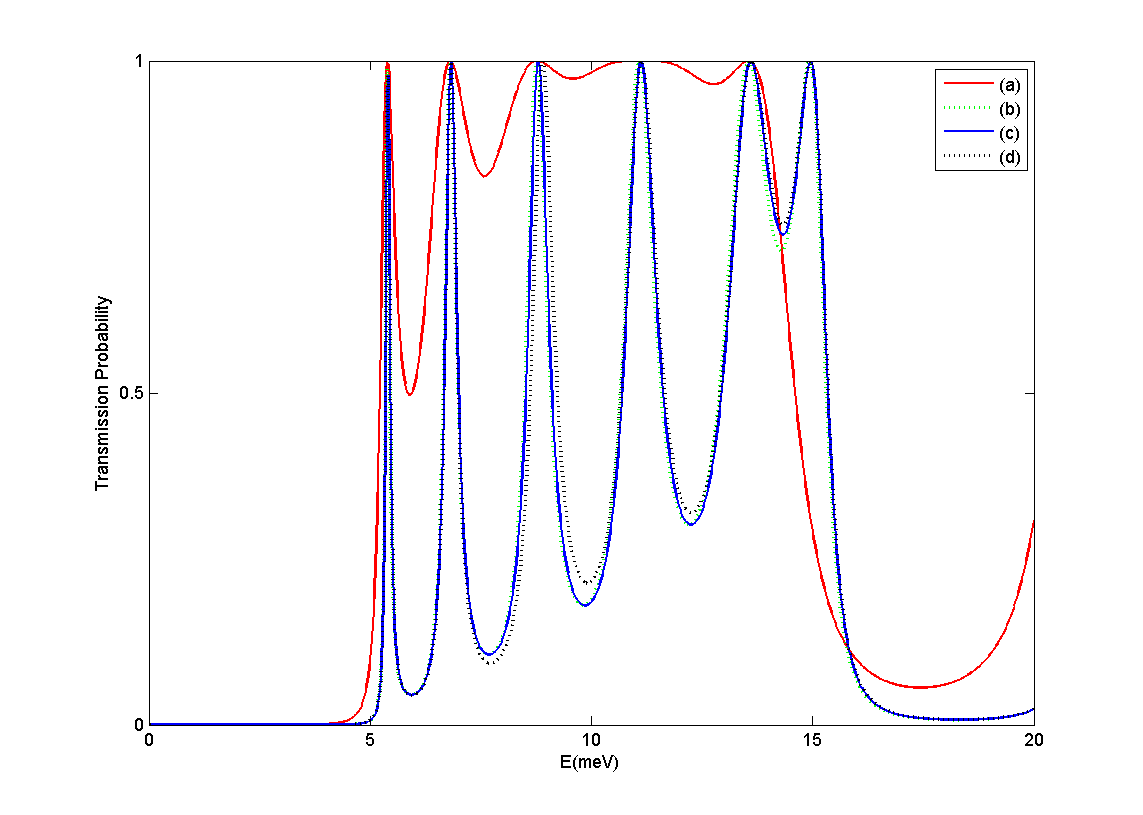}
\caption{\footnotesize (Color online) Transmission probability versus $E$ with $n=3$ and $a=14 nm$ for (a) $m^*=0.067 m_0$ for all regions, $V(x)=V_g(x)$ in $R_3$ and $V(x)=0 meV$ otherwise, (b) $m^*=0.067 m_0$ in $R_3$ and $m^*=m_0$ otherwise, $V(x)=V_g(x)$ in $R_3$ and $V(x)=0 meV$ otherwise, (c) $m^*=0.067m_0$ in $R_3$ and $m^*=m_0$ otherwise, $V(x)=V_g(x)$ in $R_3$, $V(x)=20 meV$ in $R_2$ and $R_4$, $V(x)=0 meV$ otherwise, and (d) $m^*=0.067m_0$ in $R_3$ and $m^*=m_0$ otherwise, $V(x)=V_g^{\prime}(x)$ with $h=1 nm$ in $R_3$, $V(x)=20 meV$ in $R_2$ and $R_4$, and $V(x)=0 meV$ otherwise.}\label{Fig11}
\end{figure}

In order to highlight the influences of the double barriers and the layer thickness $h$ to the transmission probability, in the case of $n=3$ and $a=14 nm$, the transmission probability as a function of $E$ is plotted in Fig.~\ref{Fig11}. Strikingly, it shows that the transmission gaps are mostly provided by the corrugations. The peaks and valleys in the tunneling domains are mainly formed by the boundaries between adjacent regions in which electron has different masses. And the boundaries can make the transmission gaps flatter. Trivially, the double barriers and the layer thickness affect on the valleys at a small scale.

\section{Conclusions}\label{5}
A particular component of the considered model is the presence of periodically corrugated thin layer. The corrugated deformation contributes the geometric potential $V_g$, a sequence attractive potential wells, to the electron across the corrugated layer. The number of wells is determined by the period number $n$ of the corrugations as $2n$. The depth of wells grows with increasing the amplitude $a$ of the corrugations. Approximately, the geometric potential can be roughly equivalent to a sequence square potential wells. The square wells can be structured by introducing periodically magnetic fields. In terms of the magnetic field, the filter designed for electron with certain energy is the vector-tunable filter~\cite{Anna2009}. By means of the surface curvature, the filter structured for electron with certain energy can be named as a curvature-tunable filter. When the layer thickness $h$ is considered, the potential $V_g$ should be replaced by $V_g^{\prime}$, which includes the modification given by the layer thickness $h$. For the sake of application, the difference between adjacent wells can be controlled by depositing the layer thickness.

The most fascinating finding in the present study is the presence of the transmission gaps and resonant tunneling domains resulting from the periodic corrugations. In the gaps electron is mostly reflected, in the tunneling domains electron readily passes. Additionally, in the resonant tunneling domains the resonant splitting peaks and valleys essentially attribute to the boundaries between adjacent regions in which electron has different effective masses, and is slightly influenced by the layer thickness $h$. These results can provide a considerable access for experimenters to fabricate nanoelectronic device. As a potential application, we can control the widths of the transmission gaps by the amplitude $a$ and period number $n$ of the corrugations, design the resonant splitting peaks and valleys by depositing different materials in adjacent regions with certain thickness. Experimentally, the nanocorrugated thin films can be obtained by the detachment method~\cite{Prinz2001, Prinz2006}. The corrugated nanofilms fabricated from narrow-gap and gapless semiconductors can be the most promising objects for experiment. In real physical experiments, Coulomb electron interaction, spin-orbital coupling, screening effects and atomic structure need to be considered for two-dimensional curved systems. These interesting questions need to be studied further.

\section*{Acknowledgments}

This work is supported by the National Natural Science Foundation of China (under Grant Nos. 11047020, 11404157, 11347126, 11304138, 11275097, 11475085, 11535005).

\end{document}